\documentclass[hyperref]{aa}  
\usepackage{graphicx}
\usepackage{txfonts}
\usepackage{hyperref}
\RequirePackage{hyperref}
\hypersetup{colorlinks=true}
\begin{document} 

\title{CRIRES-POP: a library of high resolution spectra from 1 to 5 microns}
\subtitle{II. Data reduction and the spectrum of the K giant 10 Leo}
\titlerunning{CRIRES-POP spectrum of 10 Leo}
\author{C. P. Nicholls\inst{1} \and T. Lebzelter\inst{1} \and A. Smette\inst{2} \and B. Wolff\inst{3} \and H. Hartman\inst{4,5} \and H.-U. K\"aufl\inst{3} \and N. Przybilla\inst{6} \and S. Ramsay\inst{3} \and S. Uttenthaler\inst{1} \and G. M. Wahlgren\inst{7} \and S. Bagnulo\inst{8} \and G. A. J. Hussain\inst{3,9} \and M.-F. Nieva\inst{6} \and U. Seemann\inst{10} \and A. Seifahrt\inst{11}}

\institute{Institute for Astrophysics, University of Vienna, T\"urkenschanzstrasse 17, 1180 Vienna, Austria\\
  \email{christine.nicholls@univie.ac.at}
  \and European Southern Observatory, Casilla 19001, Alonso de Cordova 3107 Vitacura, Santiago, Chile
  \and European Southern Observatory, Karl-Schwarzschild-Str. 2, 85748 Garching bei M\"unchen, Germany
  \and Material Sciences and Applied Mathematics, Malm\"o University, 20506 Malm\"o, Sweden
  \and Lund Observatory, Lund University, Box 43, 22100 Lund, Sweden
  \and Institut f\"ur Astro- und Teilchenphysik, Universit\"at Innsbruck, Technikerstr. 25/8, 6020 Innsbruck, Austria
  \and CSRA/STScI, 3700 San Martin Drive, Baltimore, MD 21218, USA
  \and Armagh Observatory, College Hill, Armagh BT61 9DG, UK
  \and Institut de Recherche en Astrophysique et Plan\'{e}tologie, Universit\'{e} de Toulouse, UPS-OMP, 31400 Toulouse, France
  \and Institut f\"ur Astrophysik, Georg-August Universit\"at G\"ottingen, Friedrich-Hund-Platz 1, 37077 G\"ottingen, Germany
  \and Department of Astronomy and Astrophysics, University of Chicago, 5640 S Ellis Avenue, Chicago, IL 60637, USA}

\date{Received ; accepted }

\abstract
{High resolution stellar spectral atlases are valuable resources to astronomy. They are rare in the $1 - 5\,\mu$m region for historical reasons, but once available, high resolution atlases in this part of the spectrum will aid the study of a wide range of astrophysical phenomena.}
{The aim of the CRIRES-POP project is to produce a high resolution near-infrared spectral library of stars across the H-R diagram. The aim of this paper is to present the fully reduced spectrum of the K giant 10 Leo that will form the basis of the first atlas within the CRIRES-POP library, to provide a full description of the data reduction processes involved, and to provide an update on the CRIRES-POP project.}
{All CRIRES-POP targets were observed with almost 200 different observational settings of CRIRES on the ESO Very Large Telescope, resulting in a basically complete coverage of its spectral range as accessible from the ground. We reduced the spectra of 10 Leo with the CRIRES pipeline, corrected the wavelength solution and removed telluric absorption with \textit{Molecfit}, then resampled the spectra to a common wavelength scale, shifted them to rest wavelengths, flux normalised, and median combined them into one final data product.}
{We present the fully reduced, high resolution, near-infrared spectrum of 10 Leo. This is also the first complete spectrum from the CRIRES instrument. The spectrum is available online.}
{The first CRIRES-POP spectrum has exceeded our quality expectations and will form the centre of a state-of-the-art stellar atlas. This first CRIRES-POP atlas will soon be available, and further atlases will follow. All CRIRES-POP data products will be freely and publicly available online.}


\maketitle

\section{Introduction}

Stellar atlases and reference spectra provide a useful observational counterpoint to laboratory- and computationally-produced atomic and molecular line lists, model stellar atmospheres, and synthetic spectra. These observational and theoretical resources are interdependent, and the research applications of both are numerous. A few uses include identification of spectral features, determination of relative abundances, estimation of stellar properties, assisting the planning of observing programs, and providing references for the search for faint companions. At very high resolution, an observational stellar atlas can provide useful feedback to line lists and model atmospheres. 

Although there are several high resolution stellar atlases available in the optical range \citep[for example the UVES-POP (Ultraviolet and Visual Echelle Spectrograph Paranal Observatory Project) library;][]{uves-pop}, there is a distinct lack of atlases in the near-infrared (NIR), especially at high resolutions (throughout this paper we use the term `near-infrared' to refer to the $1 - 5\,\mu$m region, following IPAC convention). Extant examples include the Arcturus atlas \citep{arcturus} and that of the Sun \citep{sun}, both of which were observed using Fourier Transform Spectrographs, the work of Wallace \& Hinkle and collaborators \citep{wallacehinkle96,wallacehinkle97,joyce98,wallacehinkle02}, the work of the Li\`ege group \citep[][and references therein]{zander} and, at lower resolution, the NASA Infrared Telescope Facility (IRTF) spectral library of cool stars \citep{irtflib}. The NIR range was understudied for several decades, due partly to limitations of instrumentation, and partly to telluric absorption. For several parts of the NIR, Earth's atmosphere is opaque, which defines the boundaries of atmospheric windows outside of which observing from the ground is impossible (the well-known \textit{Y, J, H, K, L}, and \textit{M} bandpasses), and telluric absorption lines are common even within these windows. In recent years, NIR instrumentation has improved significantly, and techniques for telluric line removal are much improved \citep{seifahrt,molecfit,molecfit2}. This opens up the NIR as a rich laboratory for studying cool phenomena such as stellar atmospheres, disks, exo\-planets, and circumstellar matter. Particularly for cool stars, the NIR allows us to see many spectral features whose counterparts are difficult to distinguish at optical wavelengths, including those of atomic transitions of heavy elements, and those of a variety of molecules.

The CRIRES-POP project \citep[][hereafter Paper I]{crires-pop} is in the process of producing high resolution, high signal-to-noise ratio (S/N) stellar spectral atlases that cover the entire NIR, for 26 stars that span a range of spectral types and luminosity classes. The atlases will be based on observations from the ESO/VLT high resolution NIR spectrograph, CRIRES (Cryogenic High-Resolution Infrared Echelle Spectrograph). The 26 targets were selected so the library would cover as much of the Hertzsprung-Russell diagram (HRD) as possible while satisfying the observing quality criteria. Targets were also selected to be representative of their spectral type and luminosity class, with the aim of making the atlases widely useful. An important motivation for CRIRES-POP was the provision of template spectra for the Extremely Large Telescopes, which will be most efficient in the NIR because of adaptive optics (AO) performance. The raw data, pipeline-reduced data, and final atlases will all be publicly available on the CRIRES-POP web archive and at CDS. Raw and pipeline-reduced data are already available at www.univie.ac.at/crirespop/. Each final atlas will consist of the full high resolution NIR stellar spectrum, line identifications, isotope ratios, abundances of major species, and stellar properties. 

We chose the K giant 10 Leo (HD 83240) for the first stellar atlas, partly because of our interest in cool stars and partly because it is the same spectral type as Arcturus (although with a very different metallicity), allowing us to use the Arcturus atlas as a reference when refining our data reduction process. 10 Leo is a red clump giant in the thin-disk population \citep{soubiran}, and is a long-period binary with an undetected low-mass companion. Some stellar and orbital properties of 10 Leo collected from the literature are given in Table~\ref{10leotable}. We note that the values are not all consistent, possibly due to differing analysis techniques and assumptions \citep[see for example ][]{lebzelter12}. Using the $T_{\rm{eff}}$ and distance values in the table, and assuming $A_V=0.03$ \citep{fink} and the bolometric correction from \cite{castellikurucz}, we calculate $L = 67.61\,L_{\odot}$ and $R = 11.9\,R_{\odot}$, which are clearly at odds with the literature values quoted in Table~\ref{10leotable}.

\begin{table}
\caption{Properties of 10 Leo}
\label{10leotable}
\centering
\begin{tabular}{lcc}
\hline\hline
Property & Value & Reference\\
\hline
SpT & K1 III & 1 \\
RA & 09:37:13 & 2\\
Dec & +06:50:09 & 2\\
$T_{\rm{eff}}$ & $4801\pm 89$\,K & 3\\
$L$ & 59.35\,$L_{\odot}$ & 4\\
$R$ & 14\,$R_{\odot}$ & 5\\ 
log $g$ & $2.83\pm 0.23$ & 3\\
Age & $3.51\,\pm\,1.80$\,Gyr & 6\\
Distance & 75.3\,pc & 2\\
$\rm{[Fe/H]}$ & $-0.03\pm 0.08$ & 3\\
$P_{\rm{orb}}$ & $2834\,\pm\,4$\,d & 7\\
$\gamma$ & $+20.0\,\pm\,0.1\,\rm{km\,s^{-1}}$ & 7\\
$T_{0}$ & $3\,8888\,\pm\,31$ (MJD) & 7\\
\hline
\end{tabular}
\tablebib{(1)~\cite{roman52}; (2)~\cite{hipparcos}; (3)~\cite{dasilva}; (4)~\cite{mcdonald}; (5)~\cite{pasfrac}; (6)~\cite{soubiran}; (7)~\cite{griffin}}
\end{table}

The production of a CRIRES-POP stellar spectral atlas is a two-stage process. The first stage consists of reducing the observed spectra, correcting for telluric absorption, and combining them into a single fully reduced stellar spectrum spanning $1 - 5\,\mu$m. The second stage encompasses line identification, abundance determination, and the compilation of stellar properties into the final spectral atlas. For 10 Leo, as the first published star from the CRIRES-POP library, a separate paper will be devoted to each of these two production stages. This allows us ample space to fully describe the associated data reduction and analysis, and as the third full NIR stellar spectrum and the first full CRIRES spectrum ever to be published, we feel this in-depth description is warranted. Our idea is that these two papers (comprising this paper and a forthcoming one) will serve as references for future CRIRES-POP project publications. 

The purpose of this paper is to introduce the full contents of the CRIRES-POP library now that all observations are completed, to detail the data reduction and preparation process for CRIRES-POP spectra, and to present the first reduced spectrum, that of 10 Leo. Section~\ref{observations} briefly describes the observations and summarises the properties of all stars in the library. Section~\ref{methods} describes the data reduction, telluric correction, and subsequent extensive work required for the preparation of a CRIRES-POP final spectrum. Examples of the final 10 Leo spectrum are presented in Sect.~\ref{results}. Finally, Sect.~\ref{discussion} summarises the paper, describes the next steps for the project, and explains the data that will be available and how they can be accessed.

\section{Observations}
\label{observations}

All observations for the project were made with CRIRES, the high resolution NIR echelle spectrograph at the VLT \citep{kaufl}, before it was removed for upgrade in July 2014. CRIRES had a resolving power of up to $R\,\sim100\,000$, an AO system, and wavelength coverage of $0.96 - 5.3\,\mu$m. Each observational setting, consisting of a fraction of one echelle order as isolated by the pre-disperser and intermediate slit, covered a narrow wavelength range ($\sim\,\lambda /70$), and was recorded on four detector chips, resulting in four narrow, disconnected pieces of spectrum per observational setting. These wavelength gaps between chips could be covered by observing adjacent settings, which overlapped slightly in wavelength. More detail on CRIRES is given in the CRIRES manual\footnote{At www.eso.org/sci/facilities/paranal/instruments/crires/doc.html}.

\begin{figure}
\centering
\includegraphics[width=0.5\textwidth]{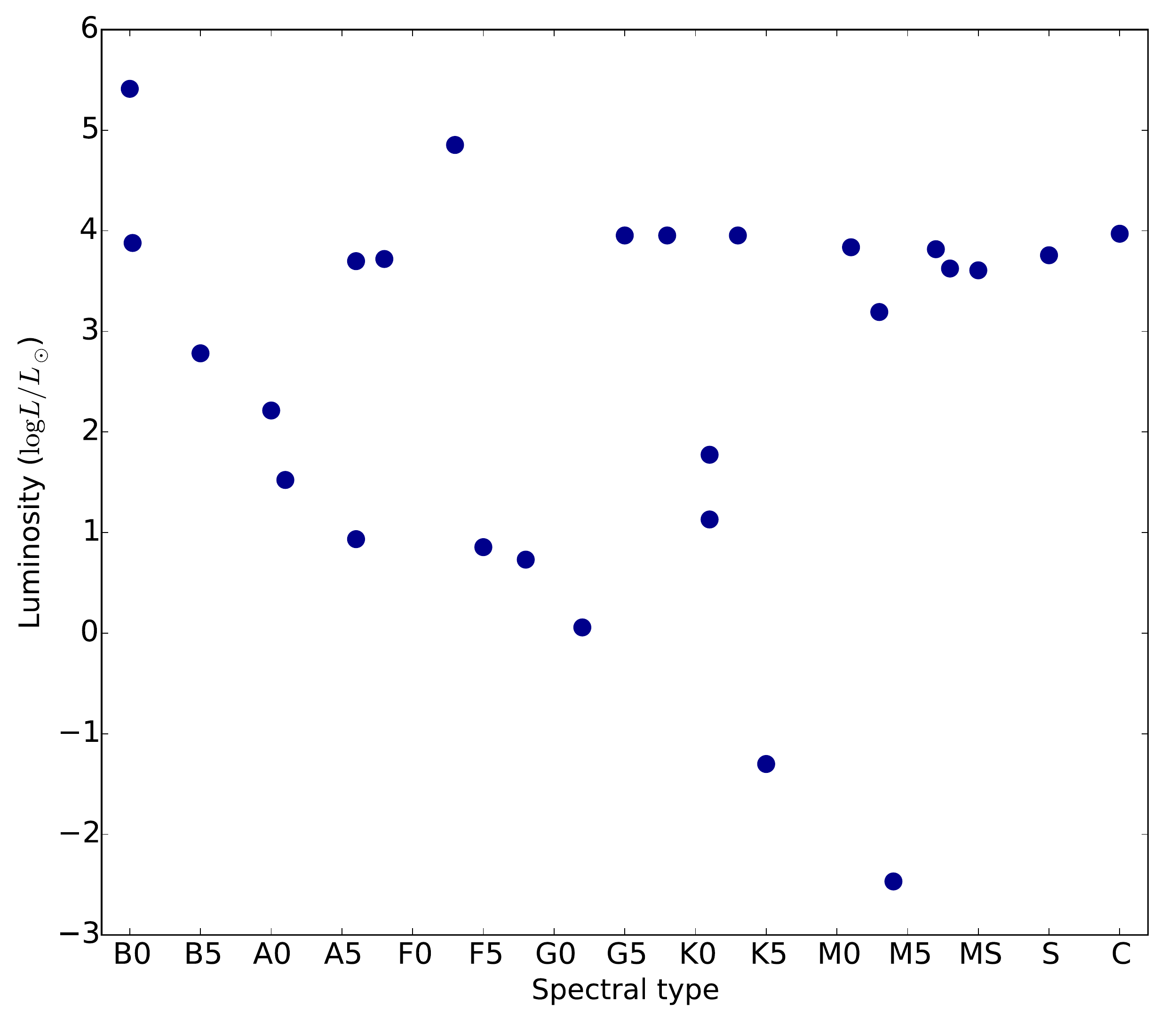}
\caption{Hertzsprung-Russell diagram of all stars in the CRIRES-POP library. Logarithm of stellar luminosity in solar units is plotted against spectral type. Luminosities were taken from the literature, primarily \cite{mcdonald}, but also \cite{anderson}, \cite{jofre}, and \cite{newton}. The data for several stars is somewhat uncertain; we found spectral type to be more homogeneous and reliable than the literature $\rm{T_{eff}}$ determinations, and some published luminosities were at odds with the accepted luminosity class. In these few cases we have moved the star to a luminosity more representative of its accepted luminosity class. Hence this diagram is intended purely for illustrative purposes, to give a visual representation of the span of the CRIRES-POP library. Included in each CRIRES-POP atlas will be updated stellar properties, which will allow us to improve this diagram as results are published.}
\label{hrd}
\end{figure}

Because the CRIRES-POP library aims to cover the entire NIR range accessible from the ground, the observing time required was considerable ($\sim$\,15 hours per star). It was therefore designed as a service mode filler program, allowing each of the 26 stars to be observed in all offered observational settings (excluding those between bandpasses) as and when possible, resulting in almost 200 observed settings per star. The target selection was influenced by both the aims of providing good-quality, widely useful final data products and by the requirements of this observing strategy, wherein the different settings for a given star might be observed several months apart. Targets were therefore selected to be bright in the NIR; representative of their spectral type and luminosity class (or abundance pattern) while also providing the library with as much breadth in spectral type and luminosity class as possible; to have little or no variability and low projected rotational velocity; and if possible, to also appear in the UVES-POP library \citep{uves-pop}. The full list of all stars in the CRIRES-POP library, along with some basic properties, is shown in Table~\ref{librarytable}. An illustrative HRD of all targets is shown in Fig.~\ref{hrd}.

By observing in almost every setting, most parts of the NIR spectrum of a star are recorded on 2 - 4 chips of different settings, which not only closes the gaps in spectral coverage between chips, but also allows an increase in S/N in the final spectrum (by a factor of up to 2) and corrects for any instrumental defects. More detail on the CRIRES-POP observing strategy and target selection can be found in Paper I.

\begin{table*}
\caption{Full library of CRIRES-POP targets, sorted by spectral type \citep[see ][for details]{crires-pop}}
\label{librarytable}
\centering
\begin{tabular}{lllccccl}
\hline\hline
HD number & Other names & Spectral type & RA (2000) & Dec (2000) & $K$ (mag) & UVES-POP? & Observing periods\\
\hline
37128 & $\epsilon$ Ori, 46 Ori, Alnilam & B0 Ia & 05:36:13 & -01:12:07 & 2.16 & n & 2011-2012 \\
149438 & $\tau$ Sco, 23 Sco & B0 V & 16:35:53 & -28:12:58 & 3.57 & n & 2010-2011 \\
120709 & 3 Cen A, V983 Cen & B5 IIIp & 13:51:50 & -32:59:39 & 4.97 & y & 2011 \\
47105 & $\gamma$ Gem, 24 Gem, Alhena & A0 IV & 06:37:43 & +16:23:57 & 1.92 & n & 2010-2011 \\
118022 & $o$ Vir, CW Vir, 78 Vir & A1p & 13:34:08 & +03:39:32 & 4.88 & y & 2010-2011 \\
73634 & e Vel & A6 II & 08:37:39 & -42:59:21 & 3.60 & y & 2009-2010 \\
39060 & $\beta$ Pic & A6 V & 05:47:17 & -51:03:59 & 3.48 & y & 2011 \\
80404 & $\iota$ Car & A8 Ib & 09:17:05 & -59:16:31 &  1.55 & y & 2011-2012 \\
74180 & b Vel & F3 Ia & 08:40:38 & -46:38:55 & 1.89 & n & 2011-2012 \\
61421 & $\alpha$ CMi, 10 CMi, Procyon & F5 IV-V & 07:39:18 & +05:13:30  & -0.65 & y & 2011-2012 \\
20010A & $\alpha$ For, LHS 1515 & F8 IV & 03:12:05 & -28:59:15 & 2.54 & y & 2009 \\
146233 & 18 Sco & G2 V & 16:15:37 & -08:22:10 & 4.19 & n & 2011-2012 \\
109379 & $\beta$ Crv, 9 Crv, Kraz & G5 II & 12:34:23 & -23:23:48 & 0.69 & y & 2010-2011 \\
99648 & $\tau$ Leo, 84 Leo & G8 Iab & 11:27:56 & +02:51:23 & 2.83 & y & 2010-2011 \\
83240 & 10 Leo & K1 III & 09:37:13 & +06:50:09 & 2.66 & y & 2009-2010 \\
138716 & 37 Lib & K1 IV & 15:34:11 & -10:03:52 & 2.25 & y & 2011 \\
225212 & 3 Cet & K3 Iab & 00:04:30 & -10:30:34 & 1.40 & y & 2010 \\
209100 & $\epsilon$ Ind & K5 V & 22:03:22 & -56:47:10 & 2.24 & y & 2011 \\
49331 & - & M1 Iab & 06:47:37 & -08:59:55 & 0.60 & y & 2010-2011 \\
224935 & YY Psc, 30 Psc & M3 III & 00:01:58 & -06:00:51 & -0.50 & y & 2009 \\
- & V2500 Oph, Barnard's star & M4 V & 17:57:48 & +04:41:36 & 4.52 & n & 2010-2011 \\
73739 & MN Vel & M7 II/III & 08:38:01 & -46:54:15 & -0.19 & n & 2011-2012 \\
18242 & R Hor & M5-8 III & 02:53:53 & -49:53:23 & -1.00 & y & 2011 \\
30959 & $o$1 Ori, 4 Ori & MS III & 04:52:32 & +14:15:02 & -0.66 & n & 2011-2012 \\
61913 & NZ Gem & S & 07:42:03 & +14:12:31 & 0.56 & y & 2010-2011 \\
134453 & X TrA & C5.5 & 15:14:19 & -70:04:46 & -0.60 & y & 2010 \\
\hline
\end{tabular}
\end{table*}

All observations for the entire CRIRES-POP library were made between 2009 October and 2012 August. We used a slit width of 0.2 arcsec, without the AO system. Nodding was done in an ABBA pattern, and integration times were selected to achieve S/N$\,\sim200$ in average seeing conditions. The standard daytime calibrations were used. Further details of the observations are given in Paper I. The raw science data are publicly available on the CRIRES-POP webpage\footnote{www.univie.ac.at/crirespop/}.

The star 10 Leo was observed between 2009 December and 2010 January. The average precipitable water vapour of the observations is between 2 and 4 mm, according to the fitted $\rm{H_{2}O}$ column from \textit{Molecfit} (see Sect.~\ref{molecfit}). The S/N of observations varied as a result of varied observing conditions. We determined S/N ratios of the observed spectra by dividing the measured flux by the flux error given by the Optimal Extraction algorithm within the CRIRES pipeline for each pixel, and computing the median over all pixels within each chip for each setting (no distinction was made between continuum and absorption lines, for practical reasons). The largest scatter in S/N is in the $H$ band, while the $K$ band spectra have the smallest S/N scatter. The mean value for the S/N ratio ranges from 241 for the $M$ band to 330 for the $YJ$ band.

\begin{figure*}
\centering
\includegraphics[width=\textwidth]{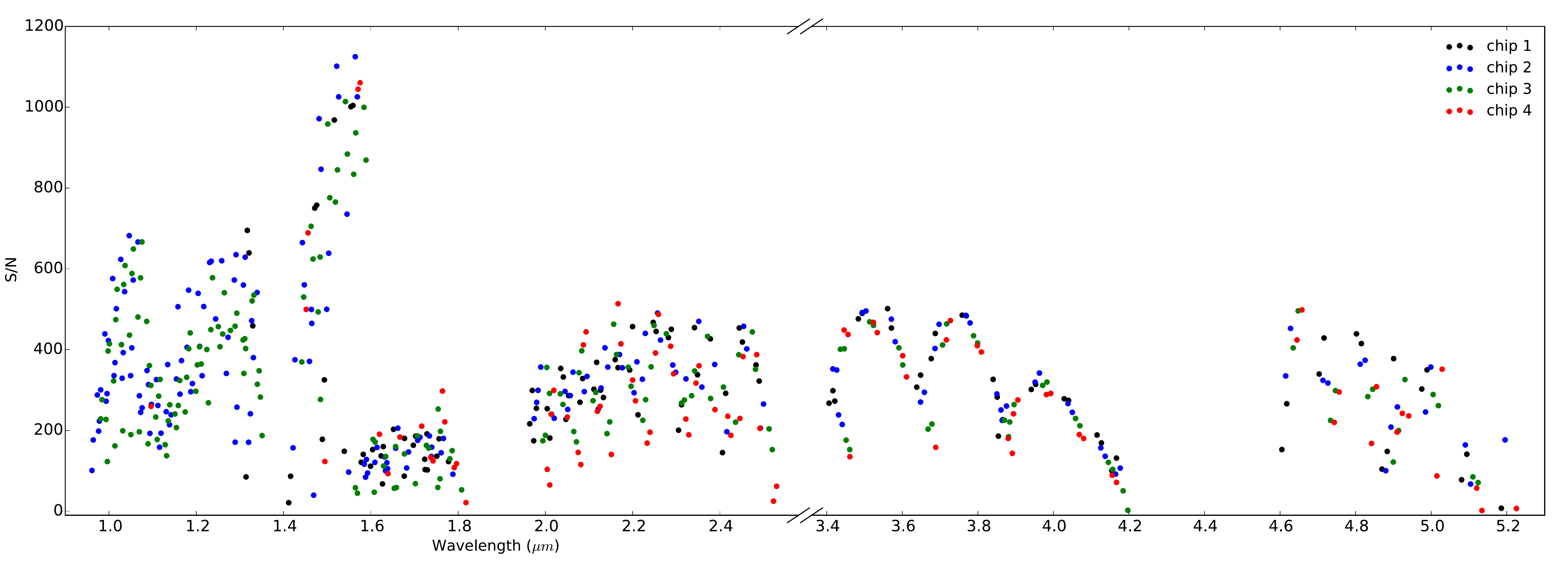}
\caption{Variation of S/N with wavelength for the 10 Leo CRIRES spectrum. One value is plotted for each chip included in the calculation of the fully reduced spectrum. See Sect.~\ref{observations} for details of how S/N was calculated, and Sect.~\ref{methods} for details of why some chips were discarded. Chip numbers within observational settings are denoted by points of different colours, as detailed in the legend.}
\label{snr}
\end{figure*}

The S/N of the entire 10 Leo CRIRES-POP spectrum is plotted against wavelength in Fig.~\ref{snr}, which shows one value for each chip, as per the calculations above. This plot does not include chips that were later discarded because of their very low S/N, contamination from adjacent orders, etc. (see Sect.~\ref{methods}). The S/N of the spectrum is generally very good; only a small minority of chips have S/N < 100. The clear structure in the $L$ band, which is also visible to some extent in the $K$ and $M$ bands, is due to the blaze function of the echelle grating: the throughput is not uniform for a given order. The large differences in S/N between the first and second halves of the $H$ band are due to two different observing dates, on which conditions were very different. The low S/N longward of $5.1\,\mu$m is due to the tapering atmospheric transmission at the edge of the $M$ band, which is also visible to some extent at the red end of the $L$ band. To our knowledge, this is the only plot showing the variation of S/N, and its dependence on both instrument characteristics and observing conditions, over the entire CRIRES wavelength range.

\section{Data reduction and preparation of the final spectrum}
\label{methods}

\subsection{CRIRES pipeline}
\label{pipeline}

Spectra of all targets were reduced with the ESO CRIRES pipeline version 2.3.2\footnote{Available at www.eso.org/pipelines}. The process included flat fielding, correcting for bad pixels and the odd-even detector readout effect, combining nod positions, extraction, and wavelength calibration. The pipeline-reduced spectra are provided as binary FITS tables and are publicly available on the CRIRES-POP webpage. More detail on the pipeline reduction process is given in Paper I. 

There are some remaining problems with the pipeline-reduced spectra, including instrumental effects we were unable to remove, and problems with the wavelength calibration. Below $2.5\,\mu$m, wavelength calibration is done using ThAr arc exposures, and above $2.5\,\mu$m, with sky lines. Because each detector chip records such a narrow wavelength range, sometimes the number of calibration lines that fall on a chip is insufficient for accurate wavelength correction. This is more prevalent for the settings calibrated with the ThAr arc lamp, and can result in spurious curvature in the wavelength solution, with wavelength inaccuracies of up to $20\,\AA$. Our procedures for correcting the wavelength calibration of affected chips are described in Sects.~\ref{molecfit} and~\ref{fiddling}.

We also encountered contamination from neighbouring orders. As mentioned in the CRIRES User Manual, at shorter wavelengths, the angular separation between adjacent orders is lower, and they can sometimes overlap, resulting in contamination of the edges of an order with features from adjacent orders. In CRIRES spectra, these effects are common on chips 1 and 4 in the $Y$ and $J$ bands (below $1.34\,\mu$m), where they affect part or sometimes all of the chip, and also occur less frequently on chips 1 and 4 in the $H$ band (from $1.34 - 1.8\,\mu$m). Often the only solution was to discard the affected chips; more detail is given in Sect.~\ref{fiddling}.

\begin{figure*}
\sidecaption
\includegraphics[width=12cm]{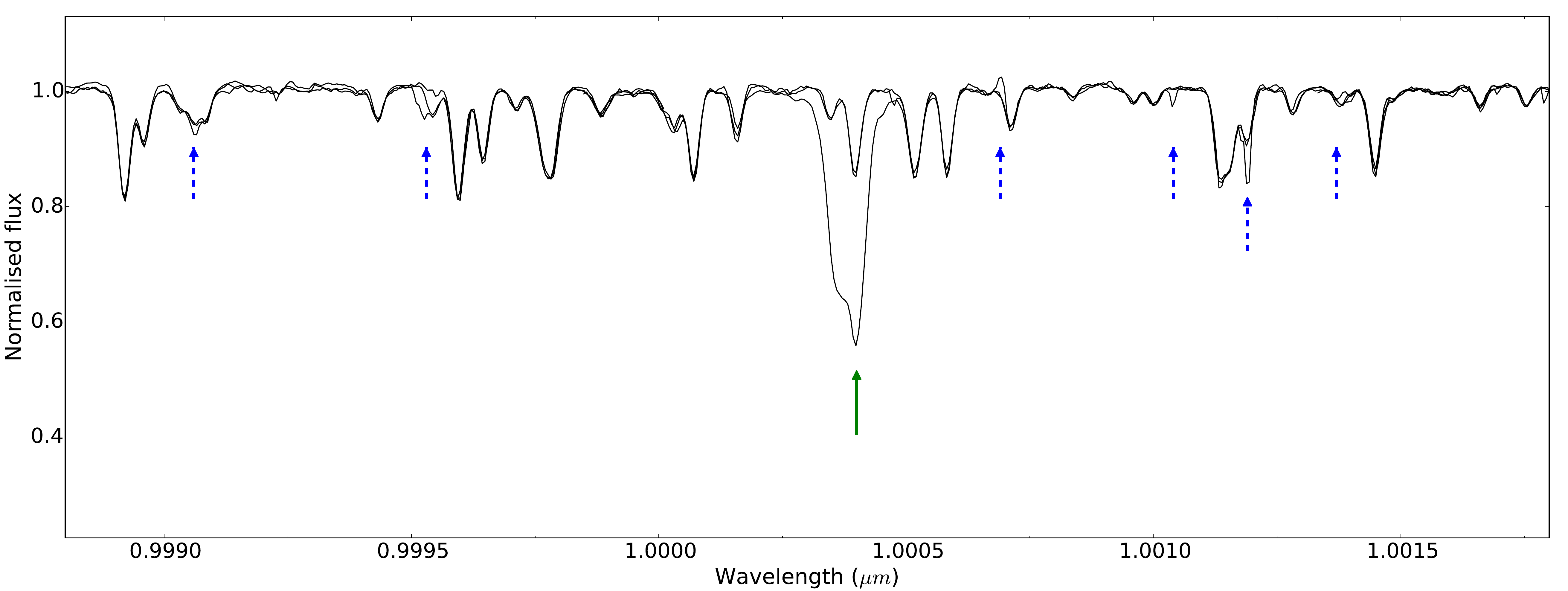}
\caption{Example of the broad absorption feature caused by the CRIRES optical ghost in the spectrum of 10 Leo. Spectra from different chips in the region of $1\,\mu$m are shown in black, demonstrating that most wavelengths are covered by 2-4 chips. This ghost is found on chip 2 of the 1005.6\,nm observational setting. The feature caused by the ghost (marked with a solid green vertical arrow) often obscures important stellar features, but these are unobscured in neighbouring settings, a fact we exploit to remove the ghost feature. Examples of small spurious lines are marked with dashed blue arrows. See Sect.~\ref{fiddling} for more detail.}
\label{ghostfig}
\end{figure*}

Finally, in settings below $\sim1.1\,\mu$m, an optical (Littrow) ghost often appears on chip 2 \citep[][see also fig.~30 in the CRIRES User Manual]{ghost}. This ghost is caused by retroreflection of light off the dispersion grating, and is probably an adjacent undispersed order. Its position on the chip, strength, and shape depend on the observational setting. It is probably due to the larger order separation at longer wavelengths that it does not appear above $1.1\,\mu$m. This ghost is not completely cancelled out by combining nod positions, and in the extracted spectrum it appears as a broad, irregularly shaped absorption feature. An example is shown in Fig.~\ref{ghostfig}. It is easy to identify but difficult to correct for; see Sect.~\ref{fiddling} for our solution.

\subsection{Telluric and wavelength correction using \textit{Molecfit}}
\label{molecfit}

The NIR is infested with telluric absorption lines, and to do science at most of these wavelengths, these must be removed. This is traditionally done by dividing the science spectrum by that of a hot `featureless' star observed close in time and airmass to the science target. However, this method is not perfect: no telluric standard star has a perfectly featureless continuum, valuable observing time is spent observing them, and nearby standard stars are not always available. For the CRIRES-POP project, observing telluric standards would have added a prohibitive amount of additional observing time. Moreover, as the stars in our library are bright, the chance that a star of appropriate type (say B0 to B5) and of similar or larger brightness is located at the same airmass as the target immediately before or after the science observation is small. In particular, in the case of 10 Leo observed at transit, the closest star of similar or smaller magnitude in the van der Bliek catalogue \citep{bliek} can only be observed with a difference of 0.4 airmass. In addition, the CRIRES-POP observations were not designed to be carried out in optimal conditions, which often meant a varying amount of precipitable water vapour during the observation of different settings within the same observing block: therefore, it is likely that observation of a telluric star would have been made with significantly different value of water vapour. Fortunately, it is now possible to instead make a model of the telluric absorption spectrum and use this to correct science spectra. This is the most suitable solution for a project such as CRIRES-POP with its large number of observations, many of which are severely affected by telluric absorption.

We chose the \textit{Molecfit} software \citep{molecfit,molecfit2} to correct for telluric absorption in the CRIRES-POP spectra. \textit{Molecfit} uses radiative transfer to model the telluric transmission spectrum at the time and location of the observations, using a combination of an atmospheric profile, atmospheric observations, and meteorological data, fitting the telluric absorption lines, stellar continuum, instrumental resolution, and wavelength dispersion of the observed spectra. Once the model is interactively and iteratively perfected, it can be used to correct the observed spectrum, producing a telluric subtraction that is at least as good as that produced in the traditional way. More details of the \textit{Molecfit} algorithm and its usage are given in \cite{molecfit}. A benefit, particularly for CRIRES spectra, is that because \textit{Molecfit} fits the wavelength dispersion, its results can be used to improve the wavelength calibration.

\begin{figure*}
\centering
\includegraphics[width=\textwidth]{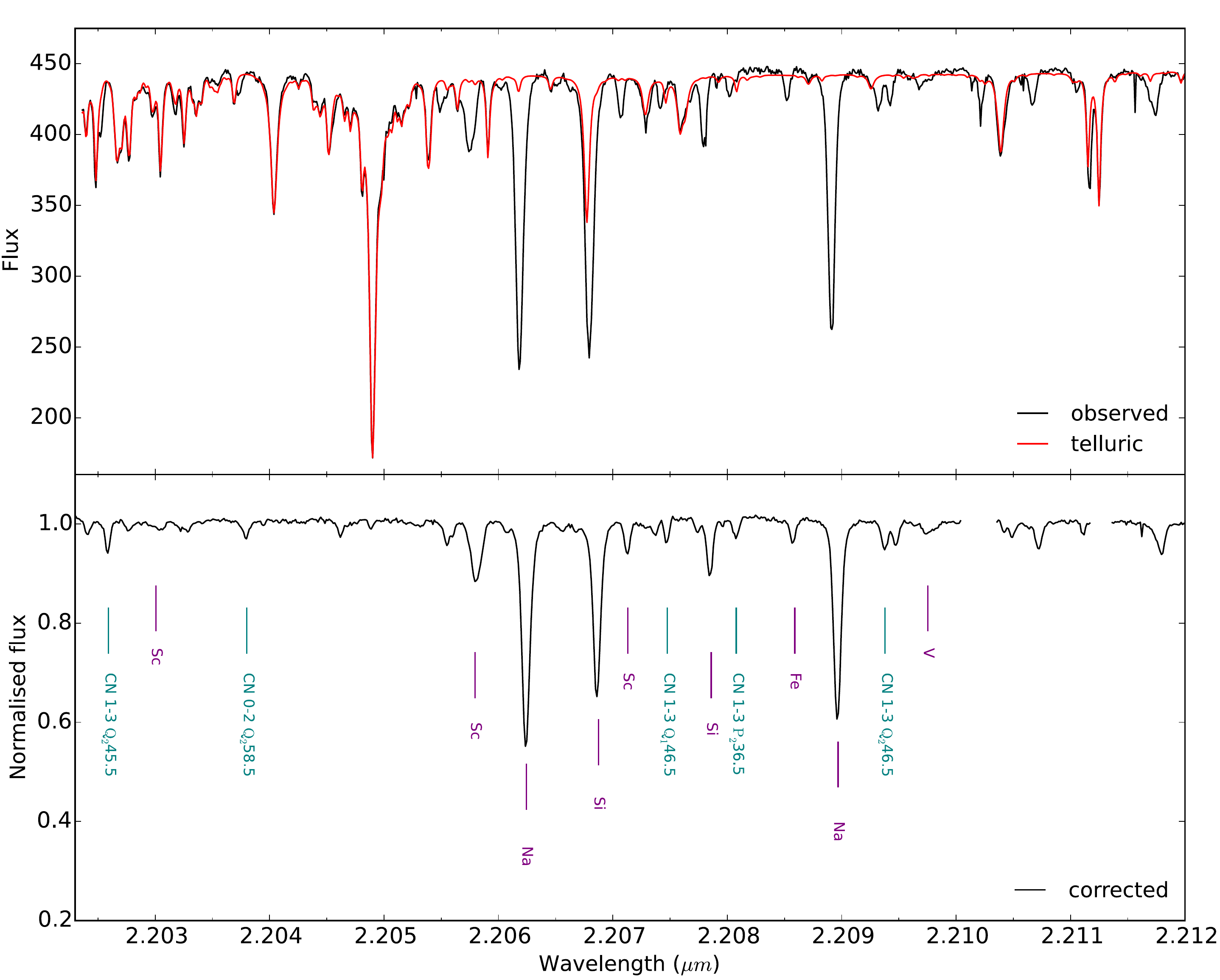}
\caption{Upper panel: Part of chip 2 of the 2219.8\,nm setting of 10 Leo, showing the pipeline-reduced stellar spectrum with telluric absorption (black line), and the telluric model made by \textit{Molecfit} (red line). Lower panel: Extract of the fully reduced 10 Leo spectrum, shown in the same wavelength region as the upper panel. Telluric lines have clearly been removed, and the spectrum has been further corrected as described in Sects.~\ref{rv} and~\ref{fiddling}. Species from the Arcturus line list are marked, with the atomic identifications in purple and the molecular identifications in teal. Several of the atomic lines in this region, including the two prominent Na lines, are magnetically sensitive. Note the subtle tellurics blended with these Na lines. See Sect.~\ref{discussion} for further details.}
\label{telluric+magnetic}
\end{figure*}

We used the GUI of version 1.0.2 of \textit{Molecfit} to correct for telluric absorption in the spectrum of 10 Leo. Initial experimentation showed that there is no loss of resolution in \textit{Molecfit}-corrected spectra by using nod-combined spectra instead of separate nodding positions. We ran \textit{Molecfit} on all four chips of an observational setting at once. We included in each fit only the telluric molecules expected for the working wavelength range (as described in the \textit{Molecfit} manual\footnote{for a complete list of telluric molecules in this wavelength range, refer to Fig.~18 of the \textit{Molecfit} manual, available at www.eso.org/pipelines/skytools/molecfit}), set the degree of the wavelength fit polynomial to 2, and used a Voigt profile approximation to fit the resolution, with a kernel varying with wavelength. We set the wavelength scale to vacuum and the instrument pixel scale to 0.086 arcsec/pix, as appropriate for CRIRES. We used wavelength exclusion masks to mask the chip edges and any purely stellar features. For cool stars like 10 Leo, the use of exclusion masks is crucial because of the number of stellar features and blends of stellar and telluric lines, which can easily confuse the telluric model, especially in regions with poor input wavelength calibration. Generally, we needed to mask any non-telluric features for best results. We used the Arcturus atlas \citep{arcturus} as a guide when determining which features were stellar and which were telluric. Once all settings were selected and the masks made, we ran the telluric fit and inspected the resulting model, comparing it to the input spectrum. We made several passes to refine the fit, tweaking the parameters and mask boundaries as necessary. Once the telluric model showed no further improvement in either the reduced $\chi ^2$ or in a visual examination of the fit, the correction was applied to that observational setting, which effectively removed the telluric lines. An example of the spectrum of 10 Leo before and after telluric correction is shown in Fig.~\ref{telluric+magnetic}, together with an example of a telluric model made by \textit{Molecfit}, .

For more than half of the observational settings, obtaining a good fit was not trivial or even not possible. These included settings with very poor wavelength calibration, settings with few or no telluric lines, settings with so many stellar features that no unblended telluric lines were available, settings with low S/N (usually $<80$, see Sect.~\ref{fiddling}), and settings with a lot of contamination from adjacent orders. We required the wavelength fit to be accurate to within one pixel. In the case of obvious wavelength calibration problems (which usually affect individual chips rather than the entire setting), we sometimes were able to fix these by reducing the convergence parameters or increasing the polynomial degree of the wavelength fit, but more often we had to fit the offending chips separately. Sometimes (most often below $2.5\,\mu$m) this still did not work and we had to manually correct the wavelength calibration later in the analysis (see Sect.~\ref{fiddling}). We also encountered wavelength calibration problems in chips with few or no telluric lines. \textit{Molecfit} uses the telluric lines to fit the wavelength dispersion, and with few lines this is difficult, and with no lines it becomes impossible to correct the wavelength at all. These chips also had their wavelength solution fixed manually later on. Where no unblended telluric lines were available, to allow \textit{Molecfit} to correctly fit both the depth of the telluric absorption features and the wavelength dispersion, we sometimes had to leave stellar-telluric blends unmasked, or mask their cores only. Chips with low S/N we sometimes had to mask mostly or entirely, and all areas of contamination had to be completely masked. Another problem we discovered occurs where there are very broad telluric bands: stray light inside CRIRES often meant that there was nonzero flux in the cores of these bands, leading to large positive residuals once telluric correction was made \citep{villa}. All such remaining problems were fixed later in the analysis (see Sect.~\ref{fiddling}).

An unfortunate result of the highly variable nature of telluric molecule abundances, the narrow wavelength ranges of CRIRES observational settings, and the inconsistent wavelength calibration from the CRIRES pipeline, is that the telluric fit calculated for a particular setting observed on a particular date cannot be applied to other settings or to the same setting in other stars. Therefore we used \textit{Molecfit} to fit and correct for telluric absorption in every observational setting of 10 Leo individually, and all observations of each star in the entire library will likely need the same treatment.

\subsection{Orbital motion of 10 Leo}
\label{rv}

To produce a final spectrum where all wavelengths are at rest, we need to measure and correct for the radial velocity in every observational setting. The sources of radial velocity shift in our spectra are the orbital motion of 10 Leo and the Earth's orbital motion.

The orbit of 10 Leo has been well studied by \cite{griffin}, who published a radial velocity curve with good phase coverage, all their radial velocity and phase data, and the orbital properties of the curve they calculated. Some of these orbital properties are reproduced in Table~\ref{10leotable}. Because we have a large number of observations, instead of measuring the radial velocity of each chip separately, we used \citeauthor{griffin}'s tabulated data to calculate the expected radial velocity of 10 Leo on each observing date (as there are far fewer observing dates than observational settings). We first calculated the orbital phase of each of our observing dates from the Modified Julian Dates (MJD) using \citeauthor{griffin}'s orbital properties, and then calculated the radial velocity corresponding to each observing date from its phase, using the orbital fit. We then assigned a radial velocity to each chip, based on its MJD. The full amplitude of the radial velocity variation of 10 Leo is $12.66\,\rm{km\,s^{-1}}$ according to \cite{griffin}, which translates into a shift of 9 pixels at $2\,\mu$m in our spectra.

We calculated the heliocentric corrections for each observing date using the ESO hourly airmass webtool\footnote{www.eso.org/sci/observing/tools/calendar/airmass.html}. A heliocentric correction was assigned to each chip based on its MJD. 

We tested our radial velocity and heliocentric velocity measurements by correcting a few sample chips for their measured orbital and heliocentric motion, and then checking whether the spectral features were at rest. All observed features were found to be at rest, which indicates that our radial velocity method was correct.

\subsection{Producing the final spectrum}
\label{fiddling}

CRIRES spectra are not provided at a constant resolving power, but at resolving powers of between $R = 90\,000$ and 100\,000, depending on the setting and its central wavelength. Additionally, as the spectrum is essentially covered by a series of narrow, disconnected pieces  (the individual chips), there is no perfect common wavelength scale. The final CRIRES-POP spectrum of 10 Leo, however, must be at a constant resolution, and to combine all the separate chips into one data product, we need to first set them on a common wavelength scale. We defined a common wavelength scale at a constant resolving power of $R = 90\,000$ (the minimum in the observed spectra, equivalent to a two-pixel resolution of $0.22\,\AA$ at $2.0\,\mu$m), then interpolated each chip's pixels to the corresponding wavelengths on the common scale. At the same time, we applied the corrections for the heliocentric motion and the radial velocity of 10 Leo (measured in Sect.~\ref{rv}), correcting the wavelengths to rest (in vacuum). Each chip was then flux normalised by dividing by the continuum fit made by \textit{Molecfit}. After inspecting the whole spectrum, which still consisted of separate chips, we noted several remaining problems that needed to be fixed before we could combine the chips into one final spectrum.

As noted in Sect.~\ref{molecfit}, even after reduction with the CRIRES pipeline and processing with \textit{Molecfit}, several chips had low S/N, incorrect wavelength solutions, or large residuals from the telluric correction. Contamination from neighbouring orders also affected several individual chips, as there is no way to remove or reduce the contamination during the data reduction process. We discarded all chips with large areas of contamination, which included most chip 1s and chip 4s in the $YJ$ band, as well as a few others across the spectrum. We also discarded any chip that had S/N $<80$ (unless its removal would leave a gap in wavelength coverage), and any chip that was mostly obscured by broad telluric absorption bands (usually occurring at the edge of bandpasses). After discarding all chips so classed as poor data, there were still many instances across the spectrum of smaller residual telluric contamination, often appearing on chips with otherwise good quality data, due to imperfect fits in \textit{Molecfit}. We opted to mask these telluric residuals, as no useable information can be extracted at the affected wavelengths, and it improves the appearance of the final spectrum, in which these masked areas appear blank. In this we followed the strategy also applied by \cite{arcturus} for telluric residuals in the Arcturus atlas. The main reason for using these masks is to avoid the appearance of a flat continuum or stellar line where we have good reason to believe the residual feature is telluric in origin. This differs from the approach of Paper I, where such areas were smoothed.

Chips with poor wavelength solutions had to be corrected manually, which proved to be time consuming. We often could not simply discard these chips as sometimes this would leave a gap in wavelength coverage, and often the wavelength problem was due to a lack of calibration lines at the data reduction stage, or telluric lines at the \textit{Molecfit} stage, which affected all chips within a certain wavelength range. We used the IRAF tasks \textit{identify} and \textit{dispcor} to correct the wavelength solution of these chips, using the Arcturus atlas line list to calibrate the wavelength dispersion using the stellar lines, and requiring the corrected wavelength to be accurate to within one pixel. In some cases, the previous wavelength solution of a chip was so inaccurate that we had to use IRAF iteratively to approach the correct solution.

Finally, we also had to manually fix the spurious absorption features left by the optical ghost (see Sect.~\ref{pipeline}), as well as several other small spurious lines, usually from bad pixels. The smaller spurious lines, which usually only covered a few pixels, were easy to fix by hand using the IRAF task \textit{splot}. However, the ghost features covered several pixels and often obscured stellar features (see Fig.~\ref{ghostfig}), so a more careful approach had to be taken. For each chip affected by the ghost, we identified another chip in a different setting with no ghost at the affected wavelengths, and replaced both the fluxes and the continuum fit of the ghost feature with those of the unaffected chip, accounting for differences in resolution (by simple interpolation), radial velocity, and heliocentric motion between the chips. 

Once we had discarded all bad chips, removed the ghosts and spurious lines, masked areas of residual telluric contamination, and corrected all poor wavelength solutions, we were able to produce the final spectrum. The velocity-corrected, continuum normalised, resampled chips were combined by taking the median flux at each wavelength in the common wavelength scale, leaving us with one final spectrum of 10 Leo.

\section{Results}
\label{results}

\begin{figure*}
\centering
\includegraphics[width=\textwidth]{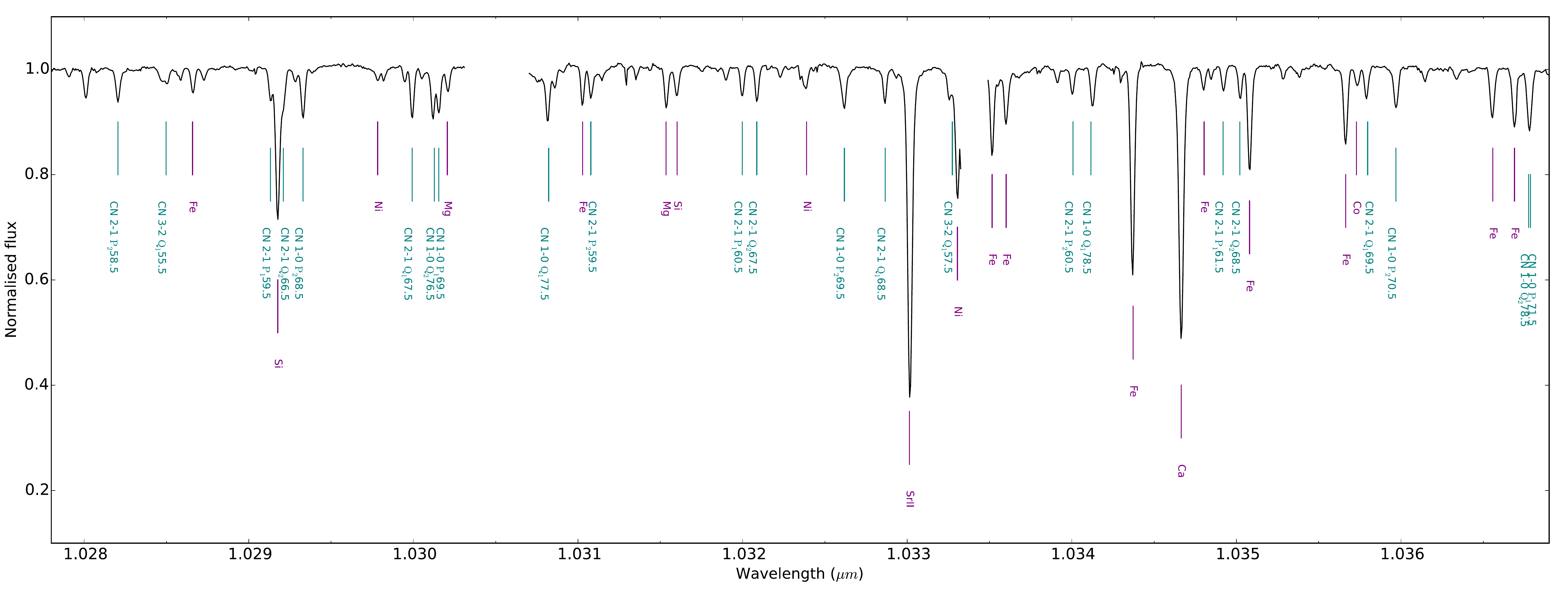}
\caption{Extract of the spectrum of 10 Leo in the $YJ$ band around the strong line of \ion{Sr}{II} at $1.033\,\mu$m, featuring several other atomic lines, and CN molecular lines. Species from the Arcturus line list are marked, with atomic identifications in purple and molecular identifications in teal.}
\label{finalspec1}
\end{figure*}

\begin{figure*}
\centering
\includegraphics[width=\textwidth]{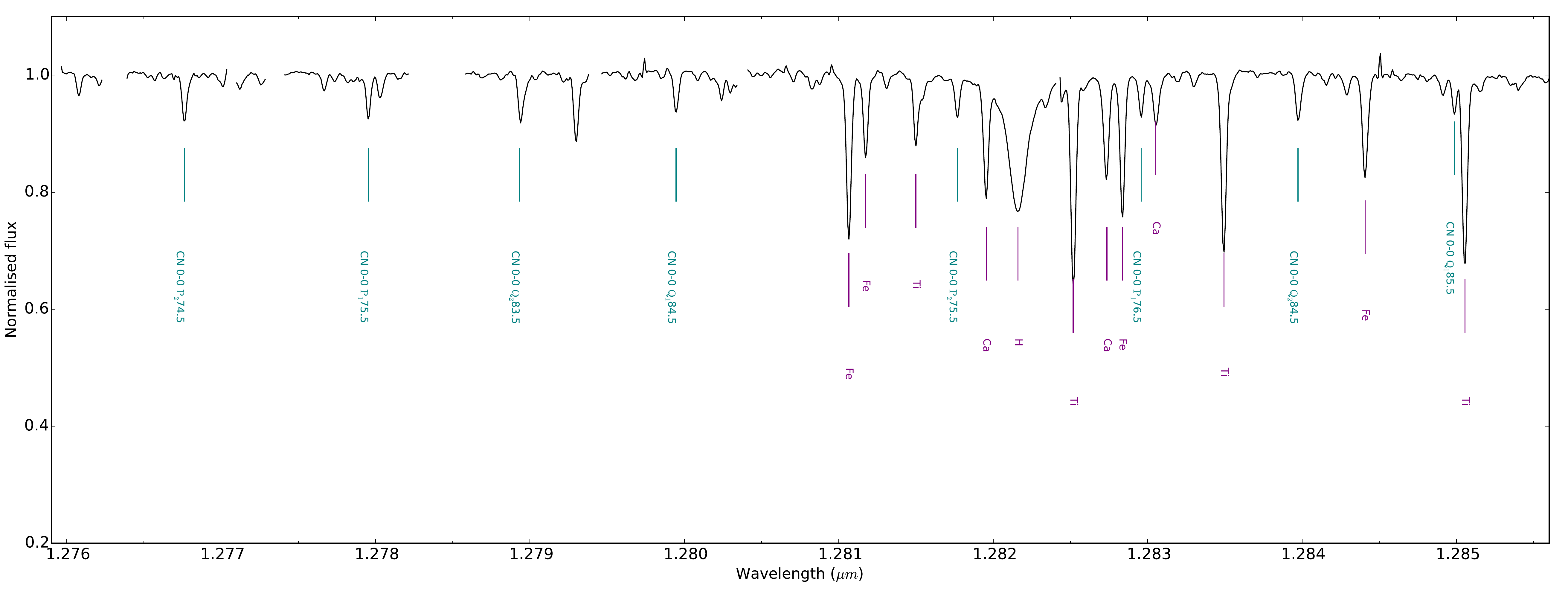}
\caption{Extract of the spectrum of 10 Leo in the $YJ$ band, including the prominent H line at $1.2822\,\mu$m, several other atomic lines, and CN molecular lines. Species from the Arcturus line list are marked as for Fig.~\ref{finalspec1}. The same wavelength range was shown in \cite{crires-pop}.}
\label{finalspec2}
\end{figure*}

\begin{figure*}
\centering
\includegraphics[width=\textwidth]{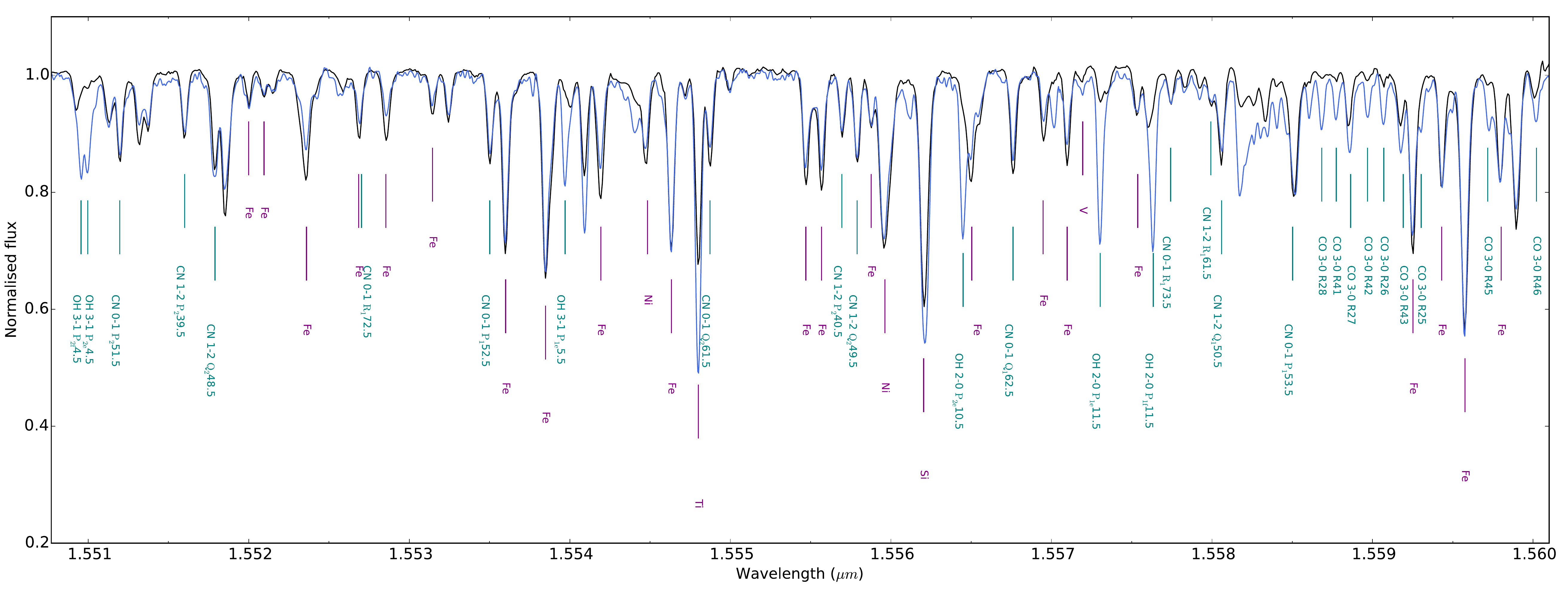}
\caption{Extract of the spectrum of 10 Leo (in black) in the $H$ band, featuring several atomic lines, and molecular lines of CN, CO, and OH. Species from the Arcturus line list are marked as for Fig.~\ref{finalspec1}. A part of the corrected Arcturus spectrum from \cite{arcturus} is overplotted (in blue), to show a sample comparison between the stars. This is discussed in Sect.~\ref{discussion}.}
\label{finalspec3}
\end{figure*}

\begin{figure*}
\centering
\includegraphics[width=\textwidth]{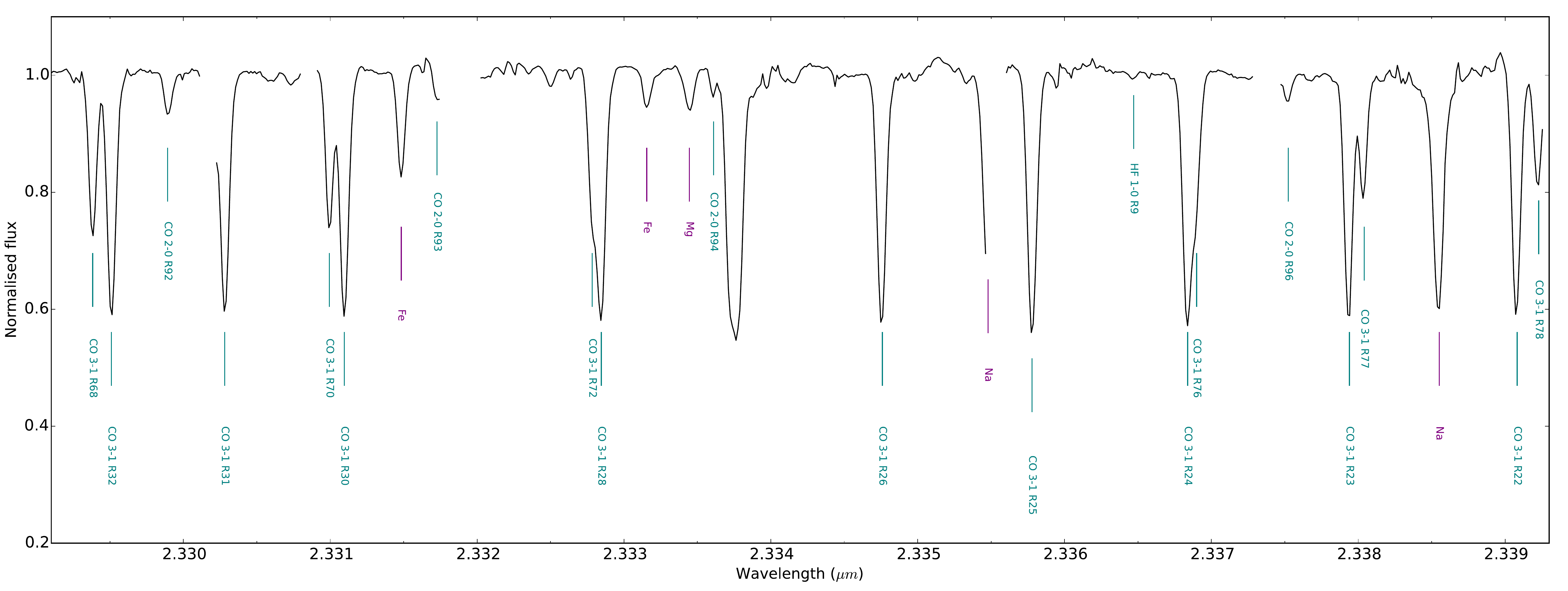}
\caption{Extract of the spectrum of 10 Leo in the $K$ band, featuring atomic lines, and molecular lines of CO and HF. Species from the Arcturus line list are marked as for Fig.~\ref{finalspec1}.}
\label{finalspec4}
\end{figure*}

\begin{figure*}
\centering
\includegraphics[width=\textwidth]{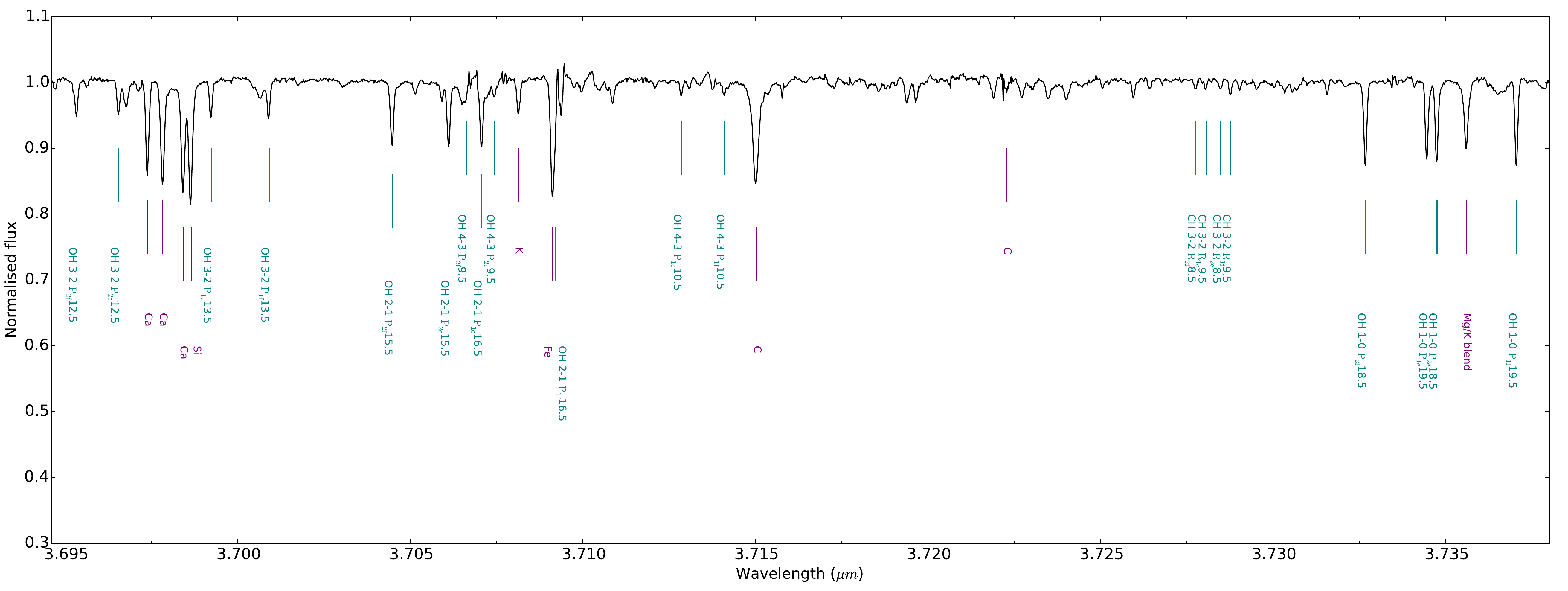}
\caption{Extract of the spectrum of 10 Leo in the $L$ band, featuring several atomic lines including C, and molecular lines of OH and CH. Species from the Arcturus line list are marked as for Fig.~\ref{finalspec1}.}
\label{finalspec5}
\end{figure*}

\begin{figure*}
\centering
\includegraphics[width=\textwidth]{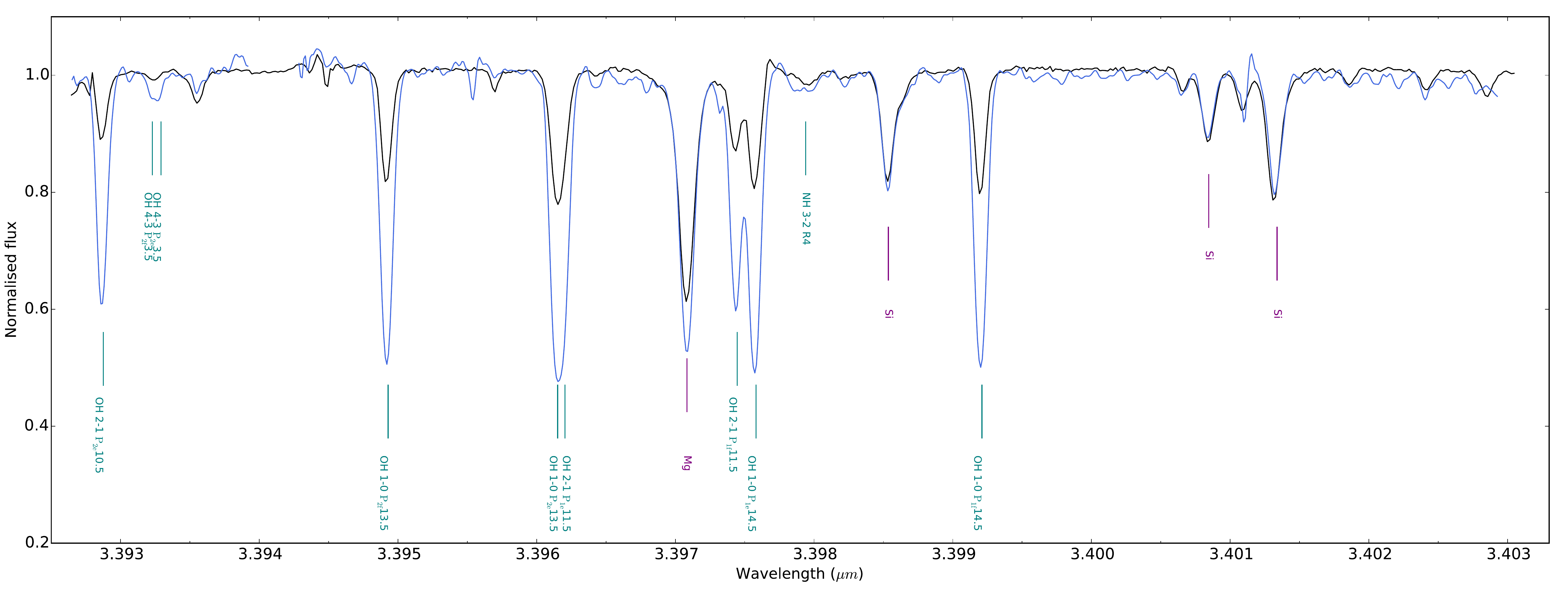}
\caption{Extract of the spectrum of 10 Leo (in black) in the $L$ band, including the prominent Mg line at $3.3971\,\mu$m, Si lines, and molecular lines of OH and NH. Species from the Arcturus line list are marked as for Fig.~\ref{finalspec1}. A part of the corrected Arcturus spectrum from \cite{arcturus} is overplotted (in blue), to show a sample comparison between the stars. This is discussed in Sect.~\ref{discussion}.}
\label{finalspec6}
\end{figure*}

\begin{figure*}
\centering
\includegraphics[width=\textwidth]{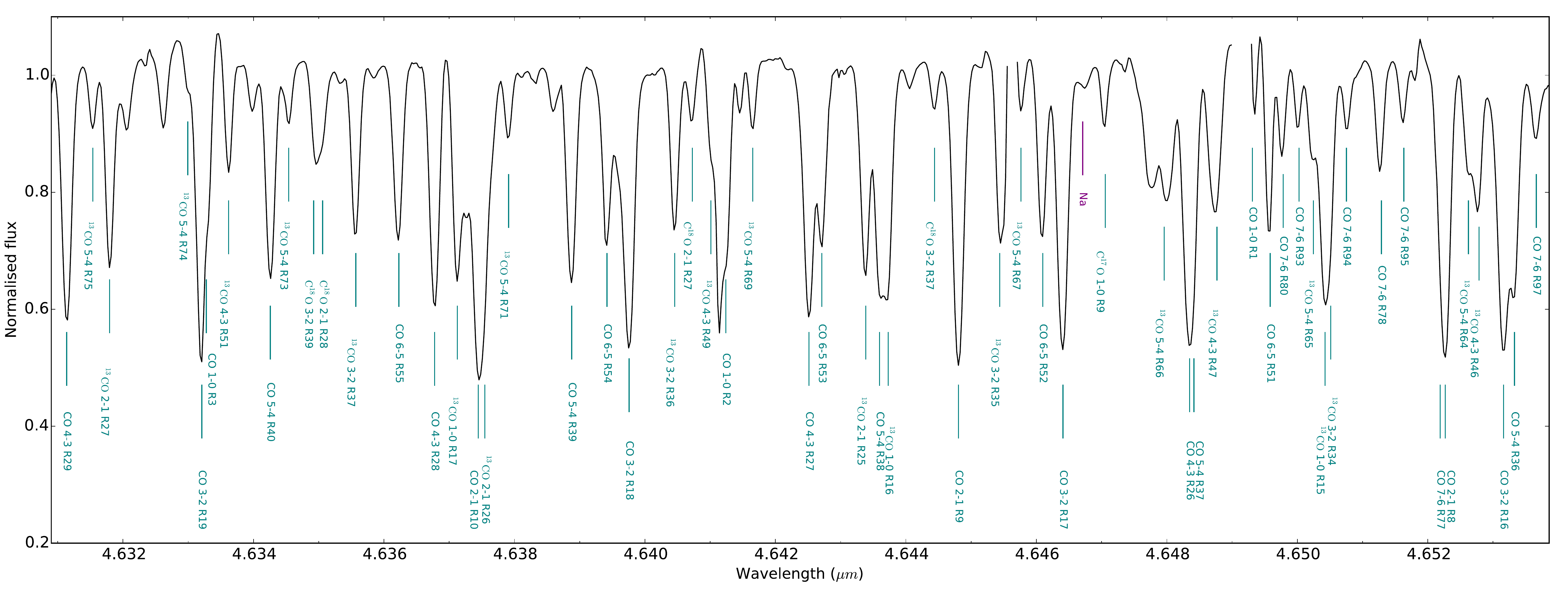}
\caption{Extract of the spectrum of 10 Leo in the $M$ band, dominated by molecular lines of $\rm{^{12}CO}$, $\rm{^{13}CO}$, $\rm{C^{17}O}$, and $\rm{C^{18}O}$, and featuring one small Na line. Species from the Arcturus line list are marked as for Fig.~\ref{finalspec1}.}
\label{finalspec7}
\end{figure*}

The final CRIRES-POP spectrum of 10 Leo covers most of the range from 0.962 to $5.245\,\mu$m at a resolving power of 90\,000. Excluded from our spectrum are regions where the Earth's atmosphere is opaque, and gaps that are due to poor coverage or discarded chips. Table~\ref{wavetable} lists the wavelength coverage of the spectrum of 10 Leo and denotes gaps (but note it does not specify the locations of the many masks). Owing to the large amount of data, it would not be informative to include a plot of the entire spectrum here, but we show in Figs.~\ref{finalspec1} to~\ref{finalspec7} some examples of the spectrum zoomed to interesting wavelength ranges. These give a sense of the S/N, resolution, frequency of masks, and overall quality of the spectrum of 10 Leo and of future CRIRES-POP data products. The spectrum has not been smoothed, and there are some areas of resulting low S/N that are due to both low S/N in some observational settings, and to fewer overlapping settings covering that area. In Figs.~\ref{finalspec3} and~\ref{finalspec6}, the corrected spectrum of Arcturus from \cite{arcturus} is overplotted on the spectrum of 10 Leo. A brief comparison of the spectra of the two stars is presented in the next section.

\begin{table}
\caption{Wavelength coverage of the CRIRES-POP spectrum of 10 Leo}
\label{wavetable}
\centering
\begin{minipage}{0.5\textwidth}
\begin{tabular}{lcl}
\hline\hline
Bandpass & Wavelength range ($\mu$m) & Reason(s) \\
\hline
$YJ$ & 0.96237 - 1.20407 & \\
 & \textit{gap} & 1\footnotetext{Reasons for gaps: (1) excluded because of contamination, (2) 1272.2\,nm setting not observed, (3) between chips of last setting in band, and (4) unresolvable wavelength problems} \\
 & 1.20420 - 1.23014 & \\
 & \textit{gap} & 1 \\
 & 1.23035 - 1.25735 & \\
 & \textit{gap} & 1; 2 \\
 & 1.25788 - 1.26390 & \\
 & \textit{gap} & 1; 2 \\
 & 1.26550 - 1.28592 & \\
 & \textit{gap} & 1 \\
 & 1.28654 - 1.30634 & \\
 & \textit{gap} & 1 \\
 & 1.30680 - 1.34453 & \\
\hline
$H$ & 1.40880 - 1.79200 & \\
 & \textit{gap} & 3 \\
 & 1.79417 - 1.80195 & \\
 & \textit{gap} & 3 \\
 & 1.80401 - 1.81140 & \\
\hline
$K$ & 1.95653 - 2.37587 & \\
 & \textit{gap} & 4 \\
 & 2.37700 - 2.52807 & \\
\hline
$L$ & 3.36694 - 4.19800 & \\
\hline
$M$ & 4.54800 - 5.15640 & \\
 & \textit{gap} & 3 \\
 & 5.16290 - 5.18720 & \\
 & \textit{gap} & 3; 4 \\
 & 5.22300 - 5.24500 & \\
\hline
\end{tabular}
\end{minipage}
\end{table}

The figures also show the species from the line list provided with the Arcturus atlas. 10 Leo is of a very similar spectral type and luminosity class as Arcturus, so this line list is a good starting point for identifying the features of 10 Leo, and we show it here for interest's sake. No corrections or further identifications have been made at this stage; all unidentified lines are presumably stellar lines. Full detailed line identifications will be published with the final atlas for 10 Leo.

The full CRIRES-POP 10 Leo spectrum is available online at the CDS\footnote{[CDS web address]}. The data are provided as five ASCII files, each containing the fully corrected spectral data for one bandpass. The file structure is simple, with Col.~1 containing the wavelength data and Col.~2 containing the flux data. The data are easy to read for both humans and computers, with all numbers printed in scientific E notation. Masks are represented with `nan' replacing the relevant fluxes, a value that plotting packages like \textit{Python matplotlib} easily ignore, allowing the user to read in and plot the entire bandpass quickly, without having to define masked regions or deal with missing data.

\section{Discussion and conclusions}
\label{discussion}

We presented the first fully reduced CRIRES-POP spectrum, that of the K1\,III giant 10 Leo, which will form the basis of the first CRIRES-POP stellar spectral atlas. This is also the first spectrum to cover (essentially) the complete spectral range of the CRIRES spectrograph. The observed CRIRES spectra were reduced with the standard pipeline, telluric- and wavelength-corrected with \textit{Molecfit}, corrected for observed radial velocity and heliocentric motion, resampled to a common wavelength scale at a constant resolving power of 90\,000, continuum normalised, and median combined to produce one final spectrum. Some poor data had to be discarded or fixed, and areas of residual telluric contamination are masked in the final spectrum. Examples of the spectrum are shown in Figs.~\ref{finalspec1} to \ref{finalspec7}. The spectrum is indicative of the quality of the CRIRES-POP library spectra as a whole, and we expect it will make an exceptional atlas.

Naturally, the spectrum has some limitations. This reference spectrum is produced from actual observations of a real star, which means that it is not perfectly smooth with absolute wavelength coverage like a synthetic spectrum would be. The S/N is not constant across the spectrum because observing conditions are variable and rarely ideal, and also because of the effect of the grating blaze function. Some unidentified telluric lines may remain in the spectrum, although we will be able to identify these as such once further spectra in the library are reduced. The continuum normalisation is not perfect everywhere, as the continua of the observed spectra are warped by broad telluric bands, influences of the spectrograph such as curvature of the spectrum across detector rows and partial polarisation of light as it passes through the slit, and the pipeline reduction process; and \textit{Molecfit} has a limited ability to fit a highly warped continuum. There are gaps due to masks, poor coverage of CRIRES settings, and wavelengths that simply cannot be observed from the ground. Some features that were visible in the Arcturus atlas are occluded by telluric lines in our spectrum, which is due to differences in radial velocity between the stars; but conversely, our spectrum includes features that were occluded for Arcturus. Some line depths are uncertain because of differing telluric subtractions on different chips.

However, an observed reference spectrum or stellar atlas has some advantages over synthetic spectra. It demonstrates to those planning observations which parts of the spectrum are scientifically interesting or useful, what S/N can be achieved, which parts of the spectrum are unusable as a result of telluric absorption, and which telluric lines can be feasibly removed without undue loss of information from the underlying stellar features. It provides feedback to telluric line lists to help improve tools like \textit{Molecfit}, and feedback to stellar line lists and atmosphere models to help improve synthetic spectra. It can be used to estimate relative abundances for other stars, study nucleosynthetic processes through isotopic ratios, and investigate stellar atmospheres and circumstellar environs. There is a clear need for both high-quality synthetic spectra and high-quality observational stellar atlases in astronomy, and each rely on the other. 

While 10 Leo is a K giant like Arcturus, significant differences in their stellar parameters can be easily spotted in a direct comparison of our CRIRES-POP spectrum with the Arcturus atlas by \cite{arcturus}. We give two examples of this comparison in Figs.~\ref{finalspec3} and~\ref{finalspec6}. The effective temperature of 10 Leo of 4801\,K (Table~\ref{10leotable}) is about 500\,K higher than that of Arcturus \citep{jonsson}. The latter also has a lower log $g$ value (1.67 vs.~2.83) and is known to be a metal-poor star, with $\rm{[Fe/H]}$ about 0.6 dex below 10 Leo. 

To have a reference for the changes produced in the spectrum by differing temperature and log $g$, we computed synthetic spectra for a COMARCS model atmosphere at 4300 and 4800\,K, and at two different log $g$ values. Details on the models and the computation of synthetic spectra can be found in \cite{aringer}.

At the higher temperature, all atomic and molecular lines in the part of the spectrum shown in Fig.~\ref{finalspec3} are weakened. In accordance with the temperature difference between the two stars, all molecular features show a lesser depth in 10 Leo than in Arcturus. This is very obvious both in the CO 3-0 lines longwards of $1.558\,\mu$m and in the OH 2-0 and
3-1 lines. For CO, the model spectra also reveal a significant sensitivity of the line depth with surface gravity, leading to a further weakening when moving from log $g$ of 1.5 to 2.5. However, the opposite trend is expected for the OH lines, which are thought to become stronger at higher log $g$, although this is not apparent in Fig.~\ref{finalspec3}. A detailed analysis of the resulting abundances of the various molecular species in 10 Leo will be presented in a forthcoming paper.

The atomic lines, in particular iron, show much less difference in strength between 10 Leo and Arcturus, despite what would be expected from the temperature difference. This supports the results from previous investigations that 10 Leo is more metal rich than Arcturus, with the higher metal abundance compensating for the reduction in line strengths by temperature.

The section of the spectrum in the $L$ band plotted in Fig.~\ref{finalspec6} is dominated by a series of fundamental OH lines. There is a very clear difference in strength between 10 Leo and Arcturus, and this agrees well with expectations from the model spectra. The Si lines in Fig.~\ref{finalspec6} show only minor differences between the two stars even though a slight temperature sensitivity is expected. As in the $H$ band, this probably reflects the higher metallicity of 10 Leo. 

We stress that these comparisons are based on a qualitative analysis using model spectra of similar temperature and surface gravity. This is done primarily for illustrative purposes. A detailed analysis of stellar parameters and abundances for 10 Leo based on the whole CRIRES-POP spectrum will be presented in a forthcoming paper.

Figure~\ref{telluric+magnetic} shows two prominent Na lines at $2.2\,\mu$m in the 10 Leo spectrum. These lines are magnetically sensitive, and their Zeeman broadening can be exploited in cool stars to measure the magnetic field strength \citep[cf.][]{johns96,shulyak}, particularly in M dwarfs. These lines are among the sparse atomic transitions in the NIR with accurate available data, whereas magnetically sensitive molecular lines in the NIR (e.g. FeH) often lack the appropriate data. Atomic lines such as the Na lines in Fig.~\ref{telluric+magnetic} serve as a benchmark when comparing field measurements from molecular and atomic lines, and are of particular relevance for later-type stars, where optical magnetic proxies become inaccessible.

The exploitation of these lines is challenged by the broadening effect of stellar rotation in later M stars and requires very accurate continuum normalisation. \cite{shulyak} has shown that without accurate telluric modelling, magnetic field measurements in early- to mid-type M stars does not yield reliable results. In the upper panel of Fig.~\ref{telluric+magnetic} we show the telluric model superimposed on the observed spectrum, which illustrates the presence of blends of telluric and stellar lines in the observed spectrum. Although the telluric blends affecting the Na lines are small, removing them is of great importance for correct stellar line fitting. Forthcoming spectral types from the CRIRES-POP library, for instance V2500 Oph (Barnard's star; M4 V), show many more weaker atomic (e.g. Na, Ti, Fe) and molecular lines (e.g. FeH). Magnetic field measurements in these lines, given the CRIRES-POP resolution, can be expected to greatly benefit from our telluric correction process.

\begin{figure}
\centering
\includegraphics[width=0.5\textwidth]{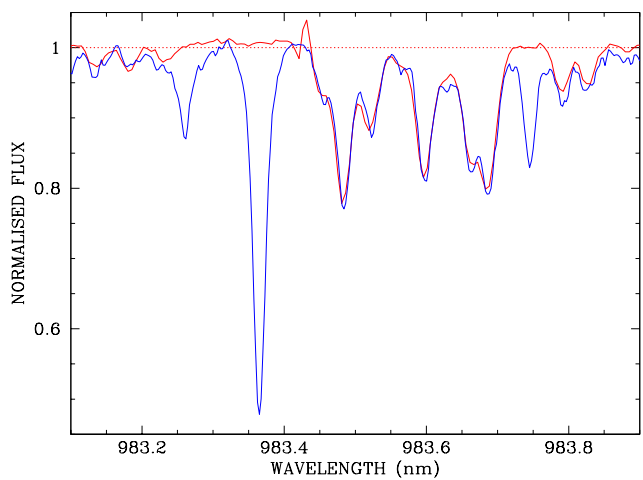}
\caption{Comparison of the UVES-POP (blue) and CRIRES-POP (red) spectra of 10 Leo in a small wavelength region of the overlap between the two spectra. The UVES-POP spectrum has not been corrected for telluric lines, and this causes the majority of the differences between the two spectra.}
\label{uvescrires}
\end{figure}

Many of the CRIRES-POP targets also appear in the UVES-POP library, meaning that these stars will have high resolution spectra with a wavelength coverage from 3000\,\AA \ to $5\,\mu$m. 10 Leo appears in both libraries, and as there is some overlap in the wavelength coverage of UVES and CRIRES, we can compare the spectra of this star taken with the two different instruments. The UVES-POP wavelength coverage ends at $1\,\mu$m, and that of CRIRES-POP starts at $0.962\,\mu$m. A small region of this overlap is shown in Fig.~\ref{uvescrires} as an example. The most striking difference is in the telluric lines, which are present in the UVES-POP spectrum and have been removed from the CRIRES-POP spectrum. The profiles of the stellar lines agree very satisfactorily, showing consistency between the two projects and the very real possibility of using the combined libraries for analysis.

Our next step is to produce the full CRIRES-POP stellar atlas of 10 Leo. In addition to the spectrum presented here, the atlas will include comprehensive line identifications, isotopic ratios, abundances of major species, and stellar properties. Some preliminary line identifications, made by overplotting the Arcturus line list on our spectrum, are shown in Figs.~\ref{finalspec1}~-~\ref{finalspec7}. The rest of the atlases for all stars in the CRIRES-POP library (see Table~\ref{librarytable}) will follow.

The central philosophy of the CRIRES-POP project is universal access to high-quality data. The entire library of atlases, including the data used to produce them, will be freely available online. The raw and pipeline-reduced spectra are available on the CRIRES-POP webpage, and the spectrum of 10 Leo presented in this paper is also now available. The final atlas data will be provided in ASCII format, and possibly in a visual format similar to the printed Arcturus atlas.

\begin{acknowledgements}
CPN acknowledges support by the Austrian Science Fund (FWF) in the form of a Meitner Fellowship under project number M1696-N27.
TL acknowledges support by the Austrian Science Fund (FWF) under project number P23737-N16. 
HH acknowledges support by the Swedish Research Council (VR) under project number 2011-4206.
M-FN acknowledges support by the Austrian Science Fund (FWF) in the form of a Meitner Fellowship under project number N1868-NBL.
Based on observations collected at the European Organisation for Astronomical Research in the Southern Hemisphere
under ESO programmes 084.D-0912(A), 085.D-0161(A), 086.D-0066(A), 087.D-0195(A), 088.D-0109(A), 088.D-0109(B), and 088.D-0109(C).
This research has made use of the SIMBAD database and the VizieR catalogue access tool, operated at CDS, Strasbourg, France.
\end{acknowledgements}

\bibliographystyle{aa} 
\bibliography{bibliographynew} 

\end{document}